\documentstyle[prl,aps,graphicx,feynmf,twocolumn]{revtex}
\newcommand{\eq}[1]{Eq. (\ref{#1})}
\newcommand{\dd}{{\rm d}}

\begin{document}
\draft
\twocolumn[\hsize\textwidth\columnwidth\hsize\csname @twocolumnfalse\endcsname

\title{Quantum critical effects on transition temperature of magnetically
mediated $p$-wave superconductivity}
\author{R. Roussev and A. J. Millis}
\address{Center for Materials Theory, Department of Physics and
Astronomy, Rutgers University, 136 Frelinghuysen Rd, Piscataway, NJ
08854}

\maketitle

\begin{abstract}
We determine the behavior of the critical temperature of magnetically
mediated $p$-wave superconductivity near a ferromagnetic quantum critical
point in three dimensions, distinguishing universal and non-universal aspects
of the result. We find that the transition temperature is non-zero at the
critical point, raising the possibility of superconductivity in the
ferromagnetic phase.
\end{abstract}
\vskip1pc]

\narrowtext
Recent experimental work has shown that superconductivity in strongly
correlated electron systems is closely associated with proximity to magnetic
quantum critical points \cite{Lonzarich98,Aeppli97,more}, suggesting it is
mediated by critical spin fluctuations \cite{Fisk98}, as indicated by
theoretical calculations \cite{Monthoux99,Mazin97}. However, the interplay of
superconductivity and criticality is not fully understood.  In this paper we
study the theoretically simplest case, namely $p$-wave superconductivity near
a ferromagnetic quantum critical point in dimension $d=3$. Our work is
complimentary to that of Abanov, Chubukov and Schmalian \cite{Chubukov00} who
studied pairing near a two dimensional antiferromagnetic quantum critical
point.

\begin{figure}
\hspace{-2ex}\includegraphics[scale=0.5]{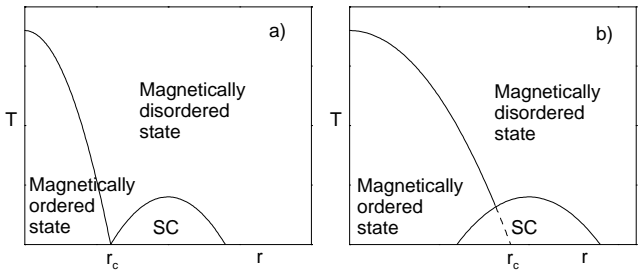}
\caption{\label{sketch}Two possible scenarios for the emergence of a
superconducting state near a quantum critical point of a magnetic system:  a)
the superconducting $T_c$ is zero in the quantum critical point (QCP), and
alternatively b) $T_c$ is finite in QCP. The parameter $r$ measures the
distance from the quantum critical point $r_c$
\protect\cite{Hertz76,Millis93}. In experimental realizations $r$ corresponds
to hydrostatic pressure \protect\cite{Lonzarich98}.}
\end{figure}

We have two motivations. One is practical: one would like to know whether the
superconducting $T_c$ vanishes as the magnetic critical point is approached
(as shown for example in the left-hand panel of Fig. \ref{sketch} and as
found by Abanov, Chubukov and Schmalian), or whether it does not (as shown in
the right-hand panel). The latter scenario raises the interesting possibility
of the coexistence of superconductivity and magnetism. This question has not
been definitively theoretically settled, because numerical difficulties have
prevented a straightforward attack \cite{Monthoux99}.

Our second motivation is theoretical. Studies of magnetically mediated
superconductivity have almost uniformly been based on the Eliashberg
equations (defined below) \cite{evenmore} which are simple generalizations of
the equations which describe conventional phonon-mediated superconductors
\cite{Scalapino}. While these equations are believed \cite{Ioffe98} to give
the leading contributions to the low-energy behavior of systems near critical
points, non-critical and high-frequency processes may also be important for
the superconducting $T_c$ (These lead, e.g. to the $\mu^*$ familiar from
conventional superconductivity) \cite{Scalapino}. Our results show how to
isolate the critical contributions and allow the magnetic analogue of $\mu^*$
to be estimated.
\begin{figure}
\hspace{-2ex}\includegraphics[]{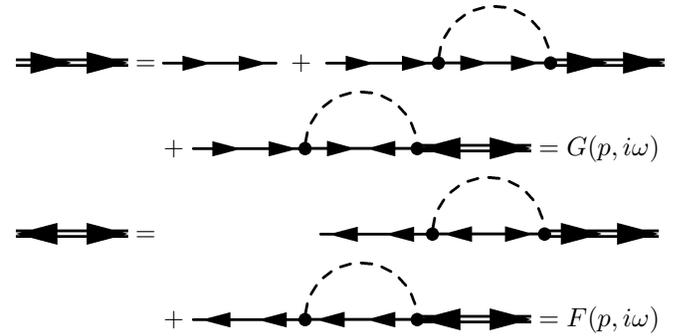}
\caption{\label{diagrams} Eliashberg equations in diagrammatic form.
$\Sigma(p,i\omega) = i\omega (1-Z_p(\omega))$ is the normal-state
quasiparticle self energy and $W(p,i\omega)$ is the anomalous self energy.}
\end{figure}

We consider a three-dimensional metal near a ferromagnetic quantum critical
point. The magnetic susceptibility is \cite{Hertz76,Millis93}
\begin{eqnarray}
& & \chi(q,\nu) = \nonumber \\
& & \left[ \displaystyle{|\nu| \over \Lambda} \cdot {q \over p_F} \tan^{-1}
\left( \displaystyle{\Lambda \over |\nu|} \cdot {p_F \over q}\right) +
\left( \displaystyle{q \over 2p_F}\right)^2 + r \right]^{-1} + \dots
\label{susceptdef}
\end{eqnarray}
where $r$ is a parameter that measures the distance of the system from its
quantum critical point (Fig. \ref{sketch}) and the ellipsis denotes less
singular terms.  Here $p_F$ is a momentum scale of the order of the Fermi
momentum and $\Lambda \sim v_F p_F$ is an energy scale of the order of the
Fermi energy.  We assume (following
\cite{Monthoux99,Mazin97,evenmore,Ioffe98}) that the coupling of these to the
electron system is given by the Eliashberg equations (shown diagrammatically
in Fig. \ref{diagrams}) for the electron self-energy
$\Sigma(p,i\omega)=i\omega(1-Z_p(\omega))$ and the anomalous self-energy
$W(p,i\omega)$.  $W(p,i\omega)$ vanishes at the superconducting critical
temperature and grows continuously in the superconducting state.  We find
$T_c$ by solving the linearized Eliashberg equations, which are
\begin{eqnarray}
i\omega(1-Z_p(\omega)) & = & g^2 s_0 \int N(\Omega_{p'}) \dd\Omega_{p'}
	\int_{-\infty}^{\infty} \dd \epsilon_{p'} \label{sigmadef} \\
	\pi T \sum_{i\omega'} & \chi & (p-p',i\omega - i\omega') {1\over
	i\omega' Z_p(\omega') - \epsilon_p'} \nonumber \\
W(p,i\omega) & = & g^2 s_l \int N(\Omega_{p'}) \dd\Omega_{p'}
	\int_{-\infty}^{\infty} \dd \epsilon_{p'} \label{wdef} \\
	\pi T \sum_{i\omega'} & \chi & (p-p',i\omega - i\omega')
	{-W(p',i\omega') \over \left[i\omega' Z_p(\omega') \right]^2 -
	\epsilon_{p'}^2}.  \nonumber
\end{eqnarray}
Here the momentum integration has been separated into integration in a
direction perpendicular to the Fermi surface ($\epsilon_p$ integration) and
integration over angles $\Omega_p$ of the spherical Fermi surface;
$N(\Omega_p)$ is the angle-dependent density of states of the quasiparticles
on the Fermi surface. The numerical factors $s_l$ relate to the nature of the
spin fluctuations and the symmetry of the pairing state.  For a system with a
Heisenberg symmetry there are three independent soft spin components all of
which contribute to $\omega Z$ so $s_0=3$. However, for spin triplet
pairing only one combination can contribute to any given component of the gap
function, so $s_1=1$. The importance of this factor was stressed by Monthoux
and Lonzarich \cite{Monthoux99}. For a system with a strong Ising anisotropy,
both $s_0$ and $s_1=1$. We will present results for the Heisenberg - Ising
crossover elsewhere.  $g$ is a constant vertex representing the interaction
between spin fluctuations and low-energy quasiparticles.  It may be
experimentally defined from the singular (as $r\rightarrow 0$) behavior of
the specific heat coefficient
\begin{equation}
\gamma = \lim_{T \rightarrow 0} {C \over T} = {m_e p_F \over 3 \hbar^3} k_B^2
Z(0).
\end{equation}

Eqs. (\ref{sigmadef}) and (\ref{wdef}) apply only for frequencies much less
than the electron bandwidth and only if the momentum dependence of $Z_p$ and
$W$ is negligible relative to the frequency dependence, conditions which are
satisfied for the leading singular behavior as $r \rightarrow 0$. We
therefore employ the Migdal approximation \cite{Migdal58} $Z_p(\omega)
\rightarrow Z(\omega)$, $W(p,\omega) \rightarrow W(\Omega_p, i\omega)$ and
perform the integral over the magnitude of the momentum.  To perform the
remaining integration over angles we note that $i\omega Z(\omega)$ has the
full symmetry of the lattice, while for $p$-wave superconductivity $W$
corresponds to the $l=1$ spherical harmonic.

\begin{figure}
\hspace{10ex}\includegraphics[scale=0.5]{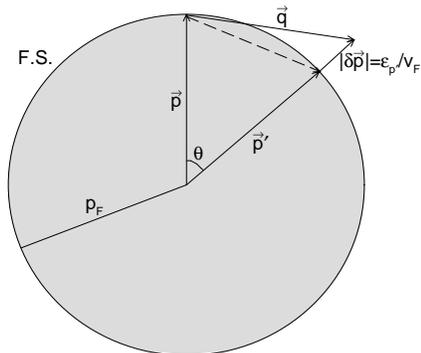}
\caption{\label{momenta}Fermi surface with momenta participating in the
interaction. The dashed line is the momentum transfer $\vec q$ when the
correction $\delta \vec p$ is neglected.}
\end{figure}

The momentum transfer $q$ carried by the spin fluctuations in \eq{susceptdef}
is given by $q^2 = (p-p')^2 = 2p_F^2 \left(1- (p \cdot p')/|p| |p'| \right) +
\epsilon_{p'}^2/v_F^2$.  The first term in $q^2$ is obtained by placing both
momenta $p$ and $p'$ on the Fermi surface while the last term is a small
correction $\delta p$ taking into account the fact that intermediate states
can explore regions close to the Fermi surface (Fig. \ref{momenta}) and will
be important as a cutoff.  We perform the $\epsilon_p$ integral, use the
angle dependences of $Z$ and $W$ and obtain
\begin{eqnarray}
|\omega| \left( 1-Z(\omega) \right) & = & - \pi T s_0 \sum_\omega
D_0(\omega - \omega') {\rm sgn}(\omega') \label{wZsimple} \\
W_l(\omega) & = & \pi T s_l \sum_\omega D_l(\omega - \omega') {W_l(\omega')
\over |\omega' Z(\omega')|}, \label{eigensimple}
\end{eqnarray}
where
\begin{equation}
D_l(\bar \nu) = 16\pi ^2 g^2 \int_0^1 N_0(x) {xP_l(1-2x^2)\dd x \over U(U
+ |\omega' Z(\omega')|/\Lambda)} \label{harmonics}
\end{equation}
with $U = \left[ (\bar \nu /x) \tan^{-1}(x/\bar \nu) + x^2 + r
\right]^{1/2}$, $\bar \nu = \nu /\Lambda$, $P_l(x)$ is a Legendre polynomial
and the $|\omega Z|/\Lambda$ comes from the $\epsilon_{p'}^2$ term.  It is
numerically very small and is important only as a cutoff at $r<T^2$ and
$\nu=0$; except in these cases we drop it. In the following we combine all
constant prefactors in $D_l$ into a single coupling constant: $16\pi^2 g^2
N_0 \rightarrow \lambda$.

To solve Eqs. (\ref{wZsimple}) - (\ref{eigensimple}) we follow Bergmann and
Rainer \cite{BergmannRainer}, defining a new order parameter $\Phi_l(\omega) =
W_l(i\omega)/|\omega Z(\omega)|$ and casting Eqs.  (\ref{wZsimple}) and
(\ref{eigensimple}) into an eigenvalue problem for an eigenvalue $\rho$
\begin{eqnarray}
\sum_{\omega'} \bigg[ s_l D_l(\omega - & \omega') & -
\left.  s_0 {|\omega' Z(\omega')| \over \pi T} \delta_{\omega\omega'}
\right] \Phi_l(\omega') = \rho \Phi_l(\omega) \label{eigeneq} \\
|\omega Z(\omega)| = & |\omega| & + \pi T \left( D_0(0) + 2
\sum_{\omega'=0}^\omega D_0(\omega - \omega') \right). \label{wZ}
\end{eqnarray}

At high temperatures the eigenvalues $\rho_n(T)$ are negative; at $T_c$ the
leading eigenvalue crosses 0. We solve the matrix system numerically; the
size of the kernel, $K_{nm} = s_l D_l(\omega_n - \omega_m) - \delta_{nm} s_0
|\omega_m Z(\omega_m)|/\pi T$ is $\sim \Lambda/(2\pi T_c)$.

For $p$-wave pairing in systems of Heisenberg symmetry the critical
temperatures are typically $\pi T_c \sim 10^{-5} \Lambda$ which translates
into numerically unmanageable kernel sizes of $N \sim 50\ 000$. We therefore
use a down-folding procedure: we separate $\Phi_l(\omega_n)$ in \eq{eigeneq}
into a low-frequency part $\Phi_l^{LOW}(\omega_n)$ with $0 \leq |n| \leq
N_{LOW}$ and high-frequency part $\Phi_l^{HIGH}(\omega_n)$ with $N_{LOW} <
|n| \leq N$.  Then \eq{eigeneq} can be written as a block linear system and
formally solved for $\Phi^{HIGH}$, yielding $K^{LOW} \cdot \Phi^{LOW} = \rho
\Phi^{LOW}$ with $K_{nm}^{LOW} = K_{nm} + \sum_{|i|, |j| > N_{LOW}} K_{ni}
\left( \rho - K_{ij} \right) ^{-1} K_{jm}$. This transformation is exact.
The simplification is that for large $N_{LOW}$ $K$ is nearly diagonal so
$K^{-1}$ may easily be computed in the ``high'' subspace. In physical terms,
this approximation retains only the diagonal scattering-dominated terms
$K_{nn} \simeq (s_1 - s_0) D_0(0) - (2n+1) (1+(2/3)\lambda s_0 + (2/3)\lambda
s_0 \ln (\Lambda / 2\pi T n)) $ and drops all off-diagonal pairing terms $s_l
D_1(\nu_{nm}) \simeq (\lambda s_1/3) \ln (\Lambda / 2\pi T|n-m|)$ for $n,m >
N_{LOW}$ in the high-frequency kernel.  We have verified that this
approximation reproduces faithfully the eigenvalues of \eq{eigeneq} for large
temperatures, and that our results are insensitive to the choice of
$N_{LOW}$.

\begin{figure}
\hspace{-3ex}
\includegraphics[scale=0.95]{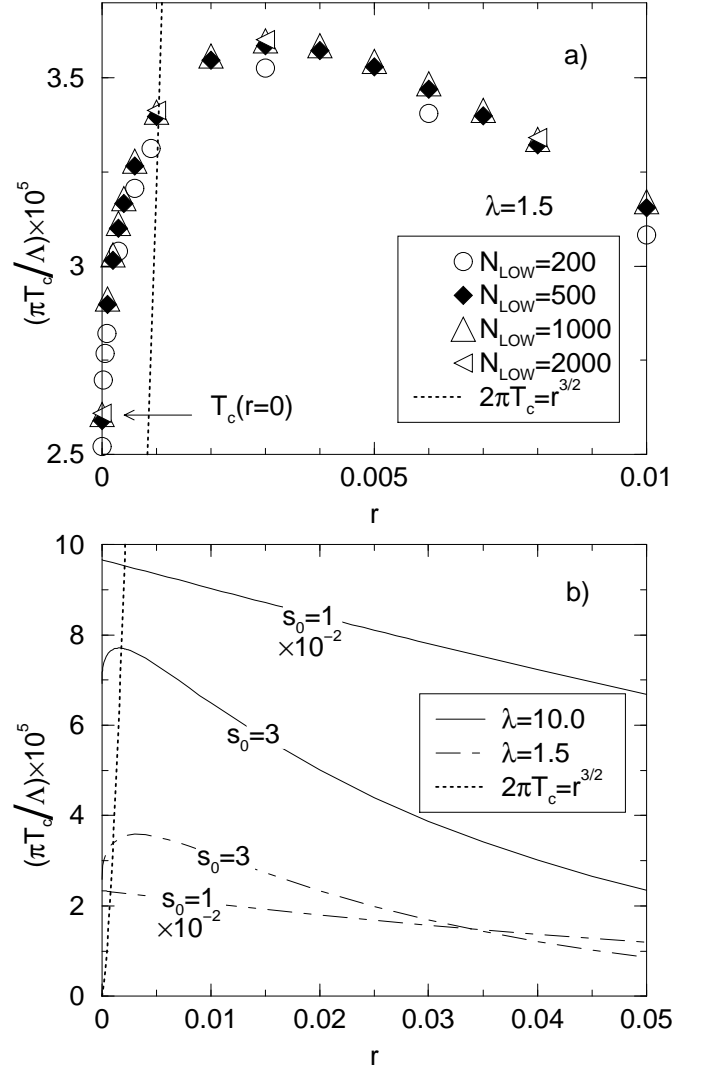}
\caption{\label{results}Results for $T_c(r)$ a) near $r=0$ for a set of
kernel sizes $N_{LOW}$ and b) in a broader range for two coupling constants
$\lambda=1.5$ and $\lambda=10.0$; in the Ising case ($s_0=1$) we have plotted
$T_c/100$ for better visual comparison with the Heisenberg ($s_0=3$) curves.
$\Lambda \sim 2p_F v_F$ is the characteristic spin-fluctuation frequency; the
curve $2\pi T_c = r^{3/2}$ separates two regimes when $T_c < r^{3/2}$ and
when $T_c > r^{3/2}$.  For the Heisenberg case ($s_0 = 3$) $N_{LOW} = 500$
while for the Ising case ($s_0 = 1$) $N = \Lambda/2 \pi T_c$.}
\end{figure}

Our results for $T_c(r)$ are shown in Fig. \ref{results}. The top panel of
the figure demonstrates the convergence of the scaling procedure with reduced
kernel size $N_{LOW}$. Kernel sizes $N_{LOW} \geq 500$ show satisfactory
convergence, so we have used sizes $N_{LOW} = 500$ in most of our work. That
large kernels are needed shows that in this problem $T_c$ is not controlled
by low-energy physics.  As previously noted \cite{Monthoux99} $T_c$ is very
low in the $p$-wave Heisenberg case because of the factor of three between
the pairing vertex and the self energy. The Ising case has not been
previously studied; we see $T_c$ is much higher.

We find that in both Heisenberg and Ising cases, $T_c(r \rightarrow 0) > 0$,
raising the interesting possibility of superconductivity extending into the
magnetic phase. We confirm that $T_c(r=0) > 0$ using a variational argument.
The ansatz $W(\omega_m) = \Delta Z(\omega_m) \Theta (\omega^{*2} -
\omega_m^2)$ allows the leading eigenvalue to be computed if $\omega^* <<
\Lambda$. At $r=0$, the leading eigenvalue becomes positive for $T < T^{var}$
with
\begin{equation}
T^{var} = {\omega^* \over \pi} \exp \left[ - { 3/2\lambda + s_0 \over s_1
\left( 1 + \ln (\Lambda / \omega^*) \right) } - {s_0 \over s_1} \right]
\label{variational}
\end{equation}
which is thus a lower bound for $T_c$. Abanov, Chubukov and Schmalian
\cite{Chubukov00}, who studied a two dimensional antiferromagnetic problem,
argued that the divergent mass enhancement associated with the critical
fluctuation would drive the superfluid stiffness and thus $T_c$ to zero. In
their case the divergent mass occurs only at one point on the Fermi surface,
so it seems to us the considerations of Hlubina and Rice \cite{Rice96} should
imply a non-zero superfluid stiffness.  In any event, in the ferromagnetic
problem of interest here the critical fluctuations are long-wavelength, and
thus do not lead to divergences in the ``transport mass'' controlling the
superfluid stiffness.

We see from Fig. \ref{results}a that the Heisenberg case $T_c(r)$ displays a
maximum at a small non-zero $r$, whereas in the Ising case there is no
maximum. We believe the small $r$ behavior is controlled by the interplay of
pairing and scattering, as discussed by Bergmann and Rainer
\cite{BergmannRainer} for $s$-wave superconductivity and by Millis, Sachdev,
Varma for $d$-wave \cite{Millis88}.  To see this mathematically we compute
$\dd T_c / \dd r$ using the Feynman-Hellman theorem \cite{BergmannRainer}:
\begin{equation}
{\dd T_c \over \dd r} = \left( \dd \rho \over \dd T_c \right) ^{-1} {\dd \rho
\over \dd r} = \left( \dd \rho \over \dd T_c \right) ^{-1} \langle \Phi |
{\dd K \over \dd r} | \Phi \rangle .
\end{equation}
The expectation value is infrared dominated numerically and we find that at
small $\omega$, $\Phi_l(\omega) \sim 1/|\omega|$. From Eqs. (\ref{eigeneq})
and (\ref{wZ}) we see there are two contributions to $\dd K / \dd r$: one
positive and proportional to $s_l$, coming from the pairing term $D_l(\nu)$
and one negative and proportional to $s_0$, from the depairing term $|\omega
Z|$ in \eq{eigeneq}. As $r \rightarrow 0$ the dominant term in $D_0$ becomes
identical to the dominant term in $D_1$. It is convenient to isolate the
contribution from zero-frequency spin fluctuations. For the leading singular
behavior in $r$ we find
\begin{eqnarray}
{\dd T_c \over \dd r} & \sim & \sum_{n=0}^\infty {s_0-s_l \over (2n+1)^2}
{\mathcal D}_0(n) - 2 \sum_{n=1}^\infty \sum_{m=0}^{n-1} \left( { - s_0\over
(2n+1)^2} \right. \nonumber \\
 & + & \left. {s_l \over (2m+1) (2n+1)} \right) {\mathcal D}_\omega(n-m)
\label{derivative}
\end{eqnarray}
Here 
\begin{equation}
{\mathcal D}_0(n) = {\lambda \over 2} {1 \over \sqrt{r}(\sqrt{r} + |\omega_n
Z|/\Lambda)}
\end{equation}
comes from differentiating \eq{harmonics} at $\bar \nu = 0$
while 
\begin{equation}
{\mathcal D}_\omega = {\lambda \over r} F \left({2\pi T \over r^{3/2}} (n-m)
\right)
\end{equation}
comes from differentiating \eq{harmonics} at $\bar \nu \neq 0$ and dropping
the $|\omega Z|$ term. The scaling function 
\begin{equation}
F(x) = \int_0^\infty {y \dd y \over (x/y + y^2 + 1)^2} ;
\end{equation}
$F(0) = 1/2$ and as $x \rightarrow \infty$, $F(x) \rightarrow (2 \pi/9
\sqrt{3}) x^{-2/3}$.

For $s_0 = 1$ (Ising case) the ${\mathcal D}_0$ term vanishes and the
${\mathcal D}_\omega$ term is negative. The pairing and depairing effects of
quasistatic ($\omega < T$) spin fluctuations exactly cancel (as in the
$s$-wave case \cite{BergmannRainer}) while at $\omega > T$ the pairing effect
wins. Thus $T_c$ monotonically increases as $r \rightarrow 0$ because the
spin fluctuations become stronger. At $r=0$, $\dd T_c / \dd r \sim
-T^{-2/3}$, i.e.  $T_c(r)$ is linear; for $r > T_c^{3/2}$ the
derivative $\rightarrow (\ln 1/r)/r$, so we expect $T_c \sim \ln^21/r$.

For $s_0=3$ (Heisenberg case) the ${\mathcal D}_0$ term is non-vanishing, and
indeed is dominant at small $r$: quasistatic spin fluctuations are
pairbreaking. At $r=0$ $T_c$ is set by the temperature at which the effect of
these fluctuations becomes small enough to allow pairing. For $r < \lambda^2
(\pi T_c)^2 \ln^2 \Lambda/T_c$, the $\omega Z$ term is important and $\dd T_c
/ \dd r \sim 1/(\sqrt{r} \lambda \pi T_c \ln \Lambda/T_c)$.  For $\lambda^2
(\pi T_c)^2 \ln^2 \Lambda/T_c < r < (\pi T_c)^{3/2}$; $\dd T_c / \dd r \sim
1/r$. In our calculations, this intermediate regime is not wide enough to
see. For larger $r$, the variation of the pairing (${\mathcal D}_\omega$)
term with $r$ becomes most important.

To summarize, we have presented a theory of the variation of a $p$-wave
superconducting $T_c$ near a ferromagnetic quantum critical point. We have
shown that the variation of $T_c$ with distance from criticality is controlled
by the low energy spin fluctuations which are theoretically tractable, and
demonstrated the crucial role played by the symmetry of the magnetic
fluctuations. We have found generically that $T_c > 0$ at the magnetic
critical point, raising the interesting possibility of superconductivity
within the ordered phase.

We acknowledge NSF DMR 9996282. A.J.M. thanks G.G. Lonzarich for many helpful
discussions.


\begin{references}
\bibitem{Lonzarich98} N. D. Mathur {\em et al.}, Nature v. {\bf 394},
39 (1998)
\bibitem{Aeppli97} G. Aeppli {\em et al.}, Science v. {\bf 278}, 1432
(1997)
\bibitem{more} Y. Maeno {\em et al.}, Nature v. {\bf 372}, 532 (1994);
D. F. Agterberg, T. M. Rice and M. Sigrist, Phys. Rev. Lett. {\bf 78}, 3374
(1997)
\bibitem{Fisk98} Z. Fisk and D. Pines, Nature v. {\bf 394}, 22 (1998)
\bibitem{Mazin97} I. I. Mazin and D. J. Singh, Phys. Rev. Lett. {\bf
79}, 733 (1997)
\bibitem{Monthoux99} P. Monthoux and G. G. Lonzarich, Phys. Rev. B {\bf
59}, 14\ 598 (1999)
\bibitem{Chubukov00} Ar. Abanov, A. V. Chubukov and J. Schmalian,
cond-mat/0005163
\bibitem{evenmore} V. J. Emery, Synth. Met. {\bf 13}, 21 (1986); K. Miyake,
S. Schmitt-Rink and C. M. Varma, Phys. Rev. B {\bf 34}, 6554 (1986); D. J.
Scalapino, E. Loh and J. E. Hirsch, Phys. Rev. B {\bf 34}, 8190 (1986); P.
Monthoux and D. Pines, Phys. Rev. B {\bf 47}, 6069 (1993)
\bibitem{Scalapino} D. J. Scalapino, in {\em Superconductivity}, edited
by R. D. Parks (Marcel Dekker, New York, 1969, Vol. 1
\bibitem{Ioffe98} see e.g. L. B. Ioffe and A. J. Millis, Usp. Fiz. Nauk {\bf
41}, 595 (1998) and references therein
\bibitem{Hertz76} J. A. Hertz, Phys. Rev. B {\bf 14}, 1165 (1976)
\bibitem{Millis93} A. J. Millis, Phys. Rev. B {\bf 48}, 7183 (1993)
\bibitem{Migdal58} A. B. Migdal, Sov. Phys. JETP {\bf 7}, 996 (1958)
\bibitem{BergmannRainer} D. J. Bergmann and D. Rainer, Z. Physik {\bf 263}, 59
(1973)
\bibitem{Rice96} R. Hlubina and T. M. Rice, Phys. Rev. Lett. {\bf 76}, 554
(1996)
\bibitem{Millis88} A. Millis, S. Sachdev and C. Varma, Phys. Rev. B {\bf
37}, 4975 (1988)
\end{references}
\end{document}